\documentclass[aps,showpacs,prx,superscriptaddress,preprintnumbers,amsmath,amssymb,twocolumn]{revtex4-2}
\usepackage[utf8]{inputenc}
\usepackage[T1]{fontenc}
\usepackage{graphicx}
\usepackage{dcolumn}
\usepackage{appendix}
\usepackage{amsmath}
\usepackage{bm}
\usepackage{amsthm}

\usepackage{braket}
\usepackage{booktabs}
\usepackage{bbold}
\usepackage{mathtools}
\usepackage{mathtools}
\usepackage{physics}
\usepackage[colorlinks]{hyperref}
\usepackage{xcolor}

\begin{document}
\allowdisplaybreaks
\title{Noise robustness of three outcome Bell certified quantum randomness}

\author{Raffaele D'Avino}

\email{raffaele.davino@icfo.eu}
\affiliation{ICFO-Institut de Ciencies Fotoniques, The Barcelona Institute of
Science and Technology, 08860 Castelldefels (Barcelona), Spain}

\author{Ignacio Perito}
\affiliation{ICFO-Institut de Ciencies Fotoniques, The Barcelona Institute of
Science and Technology, 08860 Castelldefels (Barcelona), Spain}

\author{Piotr Mironowicz}

\affiliation{Center for Theoretical Physics, Polish Academy of Sciences, al. Lotnik\'{o}w 32/46, 02-668 Warsaw, Poland}
\affiliation{Department of Algorithms and System Modeling, Faculty of Electronics, Telecommunications and Informatics, Gda\'{n}sk University of Technology, Poland}

\author{Antonio Ac\'in}

\affiliation{ICFO-Institut de Ciencies Fotoniques, The Barcelona Institute of
Science and Technology, 08860 Castelldefels (Barcelona), Spain}
\affiliation{ICREA-Institucio Catalana de Recerca i Estudis Avançats, Lluis Companys 23, 08010 Barcelona, Spain}

\author{Remigiusz Augusiak}
\affiliation{Center for Quantum-Enabled Computing, Center for Theoretical Physics,
Polish Academy of Sciences, al. Lotników 32/46, 02-668 Warsaw, Poland}

\begin{abstract}

We investigate device-independent certification of global randomness based on Bell inequality violations in bipartite scenarios with three outcomes per party. Our goal is to determine whether multi-outcome measurements allow one to surpass the amount of randomness achievable with binary outputs in realistic scenarios. We begin by analyzing several known Bell expressions and evaluating their robustness against noise for randomness certification. We then introduce a systematic method for generating new Bell expressions within structured families and perform a large-scale numerical study. We find that a substantial number of inequalities certify significant amounts of min-entropy. In particular, we identify simple inequalities that achieve near-maximal global randomness while involving a reduced number of measurement settings, thus improving the balance between certified randomness and number of inputs. Moreover, the vast majority of nontrivial certificates exhibit robustness against realistic noise, maintaining positive certified randomness away from the ideal regime. These results demonstrate that strong device-independent randomness expansion in multi-outcome scenarios is not restricted to carefully engineered inequalities, but arises generically within suitably constructed families of Bell expressions.

\end{abstract}

\maketitle

\section{Introduction}
Quantum theory allows for measurement outcomes that can be intrinsically unpredictable, even given complete knowledge of the device being used. This property has motivated the use of quantum systems for generating random numbers. A traditional approach, known as \emph{device-dependent} randomness generation, assumes that the device producing the random numbers is fully characterized. In practice, however, it is impossible to model a device perfectly, and any mismatch between the model and the actual device could compromise the security of the generated randomness~\cite{ma2016quantum}.

To address this limitation, \emph{device-independent} protocols have been introduced~\cite{mayers2003self}. These protocols rely on minimal assumptions about the devices. In particular, \emph{device-independent randomness generation} exploits the violation of a Bell inequality to certify that the outputs of uncharacterized devices are genuinely unpredictable and secure against any adversary constrained by quantum theory~\cite{pironio2010random,liu2018device}. This approach provides the strongest notion of randomness certification, as it does not depend on a detailed model of the devices’ internal workings and remains valid even in the presence of potentially malicious behavior.

Most existing protocols for device-independent randomness generation are based on two-outcome measurements in two-dimensional systems, with the CHSH \cite{clauser1969proposed} inequality serving as the canonical tool for witnessing nonlocality and quantifying extractable randomness. While conceptually simple and experimentally accessible, CHSH-based protocols fundamentally limit the amount of certifiable randomness: qubit systems constrain the structure of quantum correlations, and the binary-outcome setting bounds the entropy obtainable in a single round~\cite{mironowicz2013robustness}. These limitations motivate the exploration of higher-dimensional systems and Bell inequalities involving more outcomes, which may unlock stronger violations, richer nonlocal correlations, and ultimately more randomness per experimental use.

Recent theoretical and experimental advances have shown that high-dimensional entanglement and multi-outcome measurements \cite{collins2002bell, satwap, meyer2025robustly, buhrman2005causality} allow for more robust and flexible implementations of nonlocality. However, the implications of these advantages for device-independent randomness generation---especially under realistic noise---remain largely unexplored. In particular, while most studies have focused on the randomness obtained from the output of a single party, \emph{global randomness}, i.e., the joint unpredictability of all parties’ outcomes, may offer significantly stronger guarantees and improved rates, but its behaviour in high-dimensional noisy scenarios is not yet well understood.

In this work, we investigate the robustness to noise of device-independent randomness generation in higher-dimensional Bell scenarios, going beyond binary outcome inequalities. We consider families of Bell inequalities tailored to $d$-dimensional systems and multi-outcome measurements, and we analyse the amount of global randomness that can be certified from their quantum violations. Our goal is to determine how much noise can be tolerated before global randomness vanishes, and how this threshold depends on the system dimension, the structure of the inequality, and the underlying quantum strategy.

Our results provide several insights into the role of dimensionality in randomness generation. First, we show that higher-dimensional inequalities can certify significantly more global randomness than their qubit counterparts, even under comparable noise levels. Second, we quantify the robustness of this certification and identify regimes in which high-dimensional strategies outperform CHSH-based ones. Finally, we discuss the relevance of these findings for practical device-independent protocols, where noise resilience is a key limiting factor.

Taken together, our work highlights the potential of high-dimensional quantum systems for robust and scalable device-independent randomness generation, and it clarifies how global randomness behaves beyond the well-studied binary outcome scenario.

The organization of this paper is as follows. In sec.~\ref{sec:prel} we discuss the Bell expressions as randomness certificates. Then, in sec.~\ref{sec:RandSelected} we provide the values of the global randomness certified by selected Bell expressions. In the following sec.~\ref{sec:RandProcedure} we describe a method of parametizing Bell expression among which we perform extensive search aiming to find new Bell expressions efficient in randomness certification and we provide the results of this extensive search. We discuss the results obtained in secs~\ref{sec:RandSelected} and~\ref{sec:RandProcedure} in sec.~\ref{sec:Discussion}. We conclude in sec.~\ref{sec:Conclusion}.

\section{Preliminaries}
\label{sec:prel}

In a typical bipartite Bell experiment, two distant parties, Alice and Bob, each receive an input, $x$ and $y$, and produce corresponding outputs, $a$ and $b$, from finite sets. Repeating the experiment many times allows one to estimate the conditional probability distribution $p(a,b|x,y)$, which we assume satisfies the no-signalling condition: the marginal distribution for one party does not depend on the other party’s choice of input.

Correlations that can be explained in terms of a local-hidden-variable (LHV) model are those that can be written as
\begin{equation}
	p(a,b|x,y) = \sum_{\lambda} p(\lambda)\, p(a|x,\lambda)\, p(b|y,\lambda).
\end{equation}
Correlations that do not admit this decomposition are called nonlocal and can be witnessed by violations of Bell inequalities. These inequalities are written in terms of linear combinations of the observed probabilities,
\begin{equation}
	I(p) = \sum_{a,b,x,y} c_{abxy}\, p(a,b|x,y) \leq \beta_\mathcal{L},
\end{equation}
where $\beta_\mathcal{L}$ is the optimal value attainable by correlations compatible with LHV models. In the device-independent framework, observing a violation of this inequality provides a guarantee that the outcomes contain intrinsic randomness, independent of the devices’ internal details.

\subsection{Device-independent global randomness}

In the device-independent framework, randomness is quantified by the unpredictability of the devices’ outputs to an adversary constrained by quantum theory. In the scenario with two inputs and two outputs, a maximal violation of the CHSH inequality allows one to certify one bit of \emph{local randomness}, corresponding to the unpredictability of a single party’s output. However, within the standard CHSH scenario, this result cannot be straightforwardly extended to the certification of two bits of \emph{global randomness}, corresponding to the joint unpredictability of the outputs of all parties.

To overcome this limitation, one can consider a modified Bell inequality in which Bob is given an additional measurement input~\cite{mironowicz2013robustness}. In this construction, the CHSH inequality is augmented by the correlator $\langle A_1 B_2 \rangle$, which is maximized when Bob’s additional measurement satisfies $B_2 = A_1$. Since a maximal CHSH violation certifies that Alice’s measurements correspond to mutually unbiased bases (MUBs), measuring $A_0$ on Alice’s side and $B_2$ on Bob’s side allows one to certify two bits of global randomness. Further in the text, we denote by ($N_A$-$N_B$-$M_A$-$M_B$) scenarios where Alice (Bob) has $N_A$ ($N_B$) measurement settings with $M_A$ ($M_B$) possible outcomes. Thus CHSH belongs to ($2$-$2$-$2$-$2$) scenario, and the modified CHSH to ($2$-$3$-$2$-$2$) scenario.

\subsection{Numerical method for guessing probability}

To quantify the randomness of Alice and Bob’s joint outputs, we use the method of Nieto-Silleras \emph{et al.} introduced in \cite{nieto2014using,bancal2014more}, which characterizes the optimal guessing probability in a device-independent setting. The numerical problems formulated by this method are solved using semi-definite programming~\cite{Mironowicz_2024}.

For a fixed input pair $(x^\ast,y^\ast)$, let $(A,B)$ denote the outputs. An adversary (Eve) aims to guess $(a,b)$ using side information $E$. The guessing probability can be expressed as
\begin{equation}
    \begin{aligned}
        &P_\mathrm{guess}(AB|E,x^\ast,y^\ast)
		= \\
		&\max_{Q,\{M_{ab|z}\}} \sum_{a,b} 
		p_E(ab|z,Q)p_{AB}(ab|x^\ast,y^\ast,ab,z,Q),
	\end{aligned}
\end{equation}
where $Q$ is a quantum realization, $p_E(ab|z,Q)$ is Eve’s outcome probability, and $p_{AB}(a,b|x^\ast,y^\ast,ab,z,Q)$ is Alice and Bob’s conditional output probability.

Equivalently, one can define sub-normalised behaviours
\begin{equation}
	\tilde{p}_{ab}(a',b'|x,y) = p_E(ab|z,Q)\, p_{AB}(a',b'|x,y,ab,z,Q) \in \widetilde{\mathcal{Q}},
\end{equation}
which satisfy
\begin{equation}
	\sum_{a,b} \tilde{p}_{ab}(a',b'|x,y) = p(a',b'|x,y).
\end{equation}

In terms of these sub-normalised behaviours, the guessing probability becomes
\begin{equation}
	\begin{aligned}
	P_\text{guess}&(AB|E,x^\ast,y^\ast)
	= \max_{\{\tilde{p}_{ab}\}} \quad  \sum_{a,b} \tilde{p}_{ab}(a,b|x^\ast,y^\ast) \\
	\text{s.t.} \quad  &\sum_{a,b} \tilde{p}_{ab}(a',b'|x,y) = p(a',b'|x,y), \\
	 &\tilde{p}_{ab} \in \widetilde{\mathcal{Q}} \quad \forall a,b.
	\end{aligned}
\end{equation}

This optimisation can be relaxed via the NPA hierarchy \cite{navascues2007bounding,navascues2008convergent}, yielding an upper bound on the adversary’s guessing probability and a lower bound on the joint certifiable randomness.

\section{Randomness comparison of selected Bell expressions}
\label{sec:RandSelected}

We consider the case of local dimension $d=3$ and compare several Bell inequalities under two distinct noise models. For each inequality, we first determine its maximal quantum violation using the NPA hierarchy at level~2. Let $p_{\text{Tsir}}$ denote the (possibly unique) quantum probability distribution attaining the maximal quantum value, i.e., the Tsirelson bound, of the considered inequality. We refer to $p_{\text{Tsir}}$ as the \emph{Tsirelson distribution}.

In our protocol, randomness is certified from a distinguished pair of inputs $(x^\star,y^\star)$, referred to as the \emph{spot setting} (see, e.g.,~\cite{miller2017universal,brown2019framework,shalm2021device,liu2021device}). This setting is used in generation rounds, while the remaining inputs serve for parameter estimation. Operationally, the spot setting correspond to a pair of projective measurements that are certified to be unbiased by the Bell violation. As the shared state is maximally entangled, outcomes resulting from this pair are certified to be uniformly random.

We analyse two different noise models: in the first model, noise is introduced at the level of the quantum state via a Werner-type mixture. Let $\ket{\psi_{\text{max}}}$ denote the state achieving the maximal violation of the Bell operator. We define
\begin{equation}
	\label{eq:werner_noise}
	\rho_v = (1-v)\ket{\psi_{\text{max}}}\!\bra{\psi_{\text{max}}}
	+ v\,\frac{\openone}{d^2},
\end{equation}
where $v\in[0,1]$ and $\openone/d^2$ is the maximally mixed state on $\mathbb{C}^d\otimes\mathbb{C}^d$. At the level of probability distributions (and under the assumption of projective measurements) this corresponds to
\begin{equation}
	\label{eq:noise_distr}
	p_v = (1-v)\, p_{\text{Tsir}} + v\, p_{\text{unif}},
\end{equation}
where $p_{\text{unif}}$ is the uniform distribution over outcomes.

In the second model, noise is introduced directly at the level of the Bell expression. The observed Bell value becomes
\begin{equation}
	\label{eq:noise_ineq}
	I_v = (1-v)\, I(p_{\text{Tsir}}) + v\, I(p_{\text{unif}}),
\end{equation}
which is consistent with Eq.~\eqref{eq:noise_distr} but expressed purely in terms of the Bell value.

We compare the Salavrakos--Augusiak--Tura--Wittek--Acín--Pironio (SATWAP)~\cite{satwap} ($2$-$2$-$3$-$3$), Collins--Gisin--Linden--Massar--Popescu (CGLMP)~\cite{collins2002bell} ($2$-$2$-$3$-$3$), Buhrman--Massar (BM)~\cite{buhrman2005causality} ($3$-$3$-$3$-$3$), the inequality of Kaniewski \emph{et al.}~\cite{kaniewski2019maximal}, and that of Meyer \emph{et al.}~\cite{meyer2025robustly} (both $3$-$3$-$3$-$3$), as well as a new inequality introduced by Perito \emph{et al.} in~\cite{Barcelona} ($2$-$3$-$3$-$3$). As none of those inequalities certifies maximal randomness directly, we extend each of them by adding an extra term that requires an additional measurement on Bob's side which, together with one of Alice's measurements, certifies uniformly distributed outcomes. This modification enables global randomness certification from the designated spot setting.

\subsection{Randomness from the spot setting}

We first evaluate the amount of global randomness certified specifically from the spot setting $(x^\star,y^\star)$.

\begin{figure}[h!]
    \centering
    \includegraphics[width=0.99\linewidth]{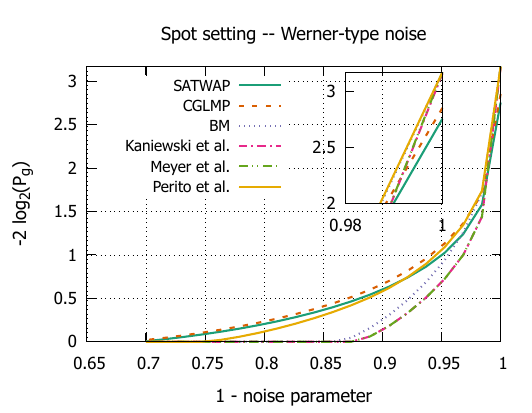}
    \caption{Certified global randomness from the spot setting as a function of the noise parameter $v$ under the Werner-type noise model of Eq.~\eqref{eq:werner_noise}. The inset highlights the region with more than two bits of certified randomness.}
    \label{fig:spot_prob_noise}
\end{figure}

Figure~\ref{fig:spot_prob_noise} shows the certified randomness as a function of the noise parameter $v$ under the Werner-type noise model defined in Eq.~\eqref{eq:werner_noise} (equivalently Eq.~\eqref{eq:noise_distr} at the level of probabilities). The BM curve overlaps with the Meyer curve and is therefore not visible.

\begin{figure}[h!]
    \centering
    \includegraphics[width=0.99\linewidth]{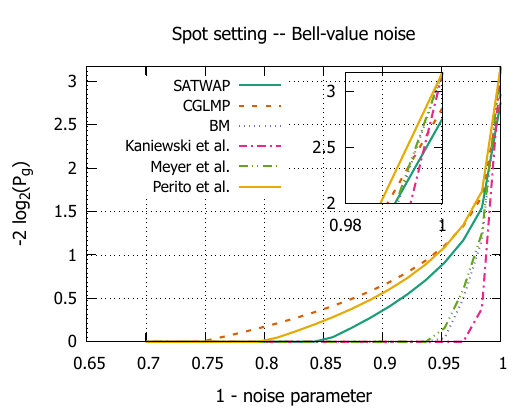}
    \caption{Certified global randomness from the spot setting as a function of the noisy Bell value $I_v$ defined in Eq.~\eqref{eq:noise_ineq}. The inset highlights the region with more than two bits of certified randomness.}
    \label{fig:spot_bell_noise}
\end{figure}

Figure~\ref{fig:spot_bell_noise} displays the certified randomness as a function of the noisy Bell value defined in Eq.~\eqref{eq:noise_ineq}, corresponding to the Bell-value noise model.

We observe, for both the Werner-type and Bell-value noise models, that the inequality introduced in~\cite{Barcelona} remains above the two-bit randomness threshold up to higher noise levels than any of the other inequalities considered. In this sense, it provides the most noise-tolerant certification of more than two bits of global randomness, surpassing the limit achievable with optimal projective qubit strategies.

\subsection{Average randomness over all inputs}

In practice, the spot setting must be selected at random among all available inputs. Consequently, the remaining settings are also implemented with nonzero probability, and it is therefore meaningful to evaluate the average randomness over all measurement inputs.

\begin{figure}[h!]
    \centering
    \includegraphics[width=0.99\linewidth]{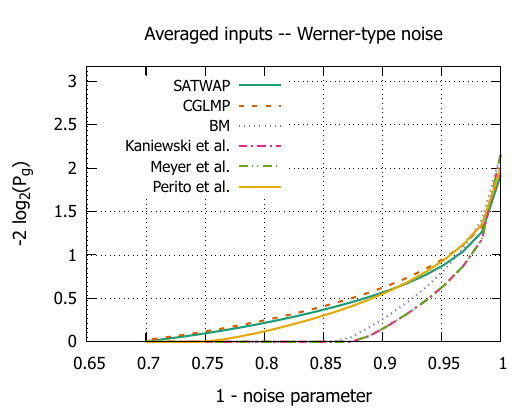}
    \caption{Average certified randomness over all inputs under the Werner-type noise model defined in Eq.~\eqref{eq:werner_noise}.}
    \label{fig:avg_prob_noise}
\end{figure}

Figure~\ref{fig:avg_prob_noise} presents the average certified randomness under the Werner-type noise model of Eq.~\eqref{eq:werner_noise}, averaged over the full set of inputs.

\begin{figure}[h!]
    \centering
    \includegraphics[width=0.99\linewidth]{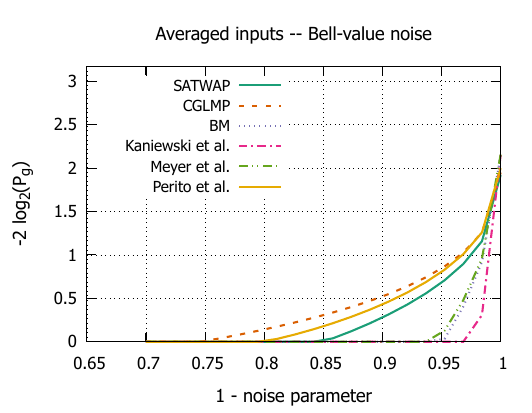}
    \caption{Average certified randomness over all inputs under the Bell-value noise model defined in Eq.~\eqref{eq:noise_ineq}.}
    \label{fig:avg_bell_noise}
\end{figure}

Figure~\ref{fig:avg_bell_noise} shows the corresponding average randomness under the Bell-value noise model of Eq.~\eqref{eq:noise_ineq}.

\section{Randomization Procedure of Bell Certificates}
\label{sec:RandProcedure}

We performed a systematic randomized search over Bell expressions with three outcomes per party. Our approach is based on fixing the measurement scenario, described by the number of measurements settings of each party, Alice and Bob, and then randomly generating Bell expressions from two distinct families, followed by a numerical evaluation of their robustness of the randomness-certification capabilities.

Throughout this section, we consider each measurement to have $d=3$ outcomes, and we reserve one specific pair of settings, $(x,y)=(0,0)$, as the \emph{spot setting}. We were considering the cases with $M$ settings of Alice and $N$ settings of Bob, for cases of $(M,N)$ equal to $(3,2)$, $(3,3)$, $(4,3)$, and $(2,4)$. In this section for brevity we write $(M,N)$ for the scenarios ($M$-$N$-$3$-$3$).

In this section we define the \emph{modulo-$d$ probability terms}
\begin{equation}
    \sum_{a+b \equiv c \;(\mathrm{mod}\;d)} P(a,b|x,y),
\end{equation}
for $c\in\{0,\ldots,d-1\}$.

\subsection{Family 1: Randomized modulo-$d$ Probability Term Selection}
\label{sec:Family1_definition}

Bell expressions belonging to \emph{family~1} are defined by assigning random modulo-$d$ probability term to all setting pairs $(x,y)\neq(0,0)$. For a scenario with $M$ settings for Alice and $N$ settings for Bob, each expression is parametrized by a vector
\begin{equation}
	\mathbf{w} = (w_1,\ldots,w_{MN-1}),
\end{equation}
whose entries are drawn independently and uniformly from the set $\{0,1,\dots,6\}$.

Each nonzero value of $w_i$ is interpreted as a pair $(k,c)$, where $k\in\{\pm1\}$ is a coefficient and $c\in\{0,1,2\}$ specifies a modulo-3 probability terms, according to
\begin{align}
	1 &\mapsto (+1,0), &
	2 &\mapsto (+1,1), &
	3 &\mapsto (+1,2), \nonumber\\
	4 &\mapsto (-1,0), &
	5 &\mapsto (-1,1), &
	6 &\mapsto (-1,2),
\end{align}
while $w_i=0$ corresponds to the absence of a term. Each $i \mapsto (x,y)$, for $i \in 1,\dots,MN-1$, and $(x,y)$ being settings of Alice and Bob different from the spot setting $(0,0)$. For each setting pair $(x,y)$, the corresponding contribution to the Bell expression is given by
\begin{equation}
	k \sum_{\substack{a,b\\ a+b \equiv c \;(\mathrm{mod}\;3)}} P(a,b|x,y),
\end{equation}
where $P(a,b|x,y)$ denotes the observed conditional probability distribution. Summing over all such terms defines a Bell expression $\mathcal{B}$.

\subsection{Family 2: Randomized Multi-Coefficient Structure}
\label{sec:Family2_definition}

The second family of Bell expressions is constructed by assigning independent coefficients to each modulo-$3$ probability term separately. In this case, each Bell expression is parametrized by a vector
\begin{equation}
	\mathbf{w} = (w_1,\ldots,w_{3(MN-1)}),
\end{equation}
with entries sampled uniformly from $\{0,1,2,3,4\}$.

Each nonzero entry $w_i$ is interpreted as a coefficient
\begin{equation}
	k \in \{+1,+2,-1,-2\},
\end{equation}
according to the mapping
\begin{align}
	1 &\mapsto +1, &
	2 &\mapsto +2, &
	3 &\mapsto -1, &
	4 &\mapsto -2,
\end{align}
while $w_i=0$ again corresponds to the absence of a term.

For each setting pair $(x,y)\neq(0,0)$ and each modulo-$d$ probability term label $c\in\{0,1,2\}$, the Bell expression includes a contribution
\begin{equation}
k \sum_{\substack{a,b\\ a+b \equiv c \;(\mathrm{mod}\;3)}} P(a,b|x,y),
\end{equation}
with independently randomized coefficients. This construction leads to a significantly larger and more flexible family of Bell expressions than in family~1.


\subsection{Family 1 statistics}

We begin with the certificates from Family~1 defined in Sec.~\ref{sec:Family1_definition}. For each scenario $(m_A,m_B) \in \{(3,2),(3,3),(4,2),(4,3)\}$ we sampled $50\,000$ random Bell expressions and evaluated the amount of certified randomness under the noise model in Eq.~\eqref{eq:noise_distr}.

Table~\ref{tab:Histogram_Family1_099999} shows the histograms for $v=0.99999$. In this nearly ideal regime, achieving certified randomness close to the theoretical maximum $-2 \frac{\log(1/3)}{\log 2} \approx 3.17$ requires four measurement settings for at least one party. In particular, the $(4,2)$ and $(4,3)$ scenarios contain the largest number of expressions certifying more than $2$ bits, with a small but non-negligible fraction approaching the maximal value (above $3$ bits). 

At the same time, the $(4,2)$ scenario is highly heterogeneous: although it contains some of the strongest certificates, the majority of sampled expressions do not certify any randomness (the dominant bin is $[0,0.001)$). This indicates that increasing the number of settings enlarges the search space substantially, most randomly chosen expressions are weak, but the tail of high-performing certificates becomes significantly richer.

For $(3,2)$ at $v=0.99999$, the situation is less favorable: this scenario has the largest number of expressions certifying essentially zero randomness and no expressions above $2.75$ bits. In contrast, $(3,3)$ already shows a substantial population of certificates above $1$ bit and a visible tail beyond $2$ bits.

The picture changes when the noise level is increased to $v=0.99$, see Table~\ref{tab:Histogram_Family1_099}. Remarkably, in every scenario there are several hundred instances (out of $50\,000$) certifying close to $2$ bits of randomness. This demonstrates that many randomly sampled expressions perform comparably to the best-known protocols based on binary outcomes. 

Importantly, the number of expressions certifying essentially zero randomness is similar to the case $v=0.99999$. Thus, most randomly generated certificates exhibit a certain degree of robustness: expressions that are non-trivial in the almost noiseless case typically remain non-trivial at $v=0.99$.

An interesting inversion occurs for $(3,2)$. While at $v=0.99999$ this scenario appeared to contain the weakest collection of certificates, at $v=0.99$ it features the largest number of instances certifying close to $2$ bits. This suggests that some expressions in this family are particularly well adapted to moderate noise, even if they are not optimal in the near-ideal regime.

\begin{table}[]
	\begin{tabular}{|l|l|l|l|l|}
		\hline
		$H_\infty$ range   & (3,2) & (3,3) & (4,2) & (4,3) \\ \hline
		{[}0,0.001)  & 46526 & 36373 & 41171 & 26297 \\ \hline
		{[}0.001,1)  & 1286  & 3937  & 2883  & 7694  \\ \hline
		{[}1,1.5)    & 0     & 2930  & 2032  & 6803  \\ \hline
		{[}1.5,2)    & 1639  & 4080  & 3327  & 5932  \\ \hline
		{[}2,2.5)    & 0     & 1642  & 25    & 2317  \\ \hline
		{[}2.5,2.75) & 549   & 1038  & 534   & 943   \\ \hline
		{[}2.75,3)   & 0     & 0     & 0     & 11    \\ \hline
		{[}3,3.1)    & 0     & 0     & 0     & 2     \\ \hline
		{[}3.1,3.17) & 0     & 0     & 28    & 1     \\ \hline
	\end{tabular}
	\caption{Histogram of certified randomness for 50000 expressions from Family 1 at noise level $0.99999$.}
	\label{tab:Histogram_Family1_099999}
\end{table}

\begin{table}[]
	\begin{tabular}{|l|l|l|l|l|}
		\hline
		$H_\infty$ range   & (3,2) & (3,3) & (4,2) & (4,3) \\ \hline
		{[}0,0.001)  & 46526 & 36931 & 41171 & 28099 \\ \hline
		{[}0.001,1)  & 1286  & 7996  & 6118  & 16918 \\ \hline
		{[}1,1.5)    & 1639  & 4731  & 2326  & 4761  \\ \hline
		{[}1.5,2)    & 549   & 342   & 385   & 222   \\ \hline
		{[}2,2.5)    & 0     & 0     & 0     & 0     \\ \hline
		{[}2.5,2.75) & 0     & 0     & 0     & 0     \\ \hline
		{[}2.75,3)   & 0     & 0     & 0     & 0     \\ \hline
		{[}3,3.1)    & 0     & 0     & 0     & 0     \\ \hline
		{[}3.1,3.17) & 0     & 0     & 0     & 0     \\ \hline
	\end{tabular}
	\caption{Histogram of certified randomness for 50000 expressions from Family 1 at noise level $0.99$.}
	\label{tab:Histogram_Family1_099}
\end{table}

\subsection{Family 2 statistics}

We now turn to Family~2, defined in Sec.~\ref{sec:Family2_definition}. Overall, this family appears less effective in certifying high amounts of randomness.

For $v=0.99999$ (Table~\ref{tab:Histogram_Family2_099999}), no sampled expression certifies randomness close to the theoretical maximum. Although there are instances exceeding $2$ bits in all scenarios, their number is significantly smaller than in Family~1, and the extreme high-value bins are essentially empty.

For $v=0.99$ (Table~\ref{tab:Histogram_Family2_099}), the distributions contract further: no expression certifies more than $2$ bits in any scenario. Nevertheless, as in Family~1, the number of expressions certifying zero randomness remains comparable to the near-ideal case. This again indicates a certain structural robustness of the sampled certificates: moderate noise reduces the achievable randomness but does not dramatically increase the fraction of completely useless expressions.

Comparing the two families, Family~1 clearly has a heavier high-randomness tail, especially in scenarios with four settings. Family~2, while still capable of producing non-trivial certificates, appears statistically less likely to generate near-optimal ones under random sampling. This suggests that the structural constraints defining Family~2 restrict the space of potentially strong Bell expressions more severely.

\begin{table}[]
	\begin{tabular}{|l|l|l|l|l|}
		\hline
		$H_\infty$ range   & (3,2) & (3,3) & (4,2) & (4,3) \\ \hline
		{[}0,0.001)  & 41335 & 32573 & 36643 & 27901 \\ \hline
		{[}0.001,1)  & 2884  & 5326  & 4715  & 6797  \\ \hline
		{[}1,1.5)    & 1908  & 3881  & 2744  & 5122  \\ \hline
		{[}1.5,2)    & 1949  & 4383  & 3431  & 6102  \\ \hline
		{[}2,2.5)    & 1273  & 2951  & 1957  & 3366  \\ \hline
		{[}2.5,2.75) & 644   & 877   & 507   & 710   \\ \hline
		{[}2.75,3)   & 7     & 9     & 3     & 2     \\ \hline
		{[}3,3.1)    & 0     & 0     & 0     & 0     \\ \hline
		{[}3.1,3.17) & 0     & 0     & 0     & 0     \\ \hline
	\end{tabular}
	\caption{Histogram of certified randomness for 50000 expressions from Family 2 at noise level $0.99999$.}
	\label{tab:Histogram_Family2_099999}
\end{table}

\begin{table}[]
	\begin{tabular}{|l|l|l|l|l|}
		\hline
		$H_\infty$ range   & (3,2) & (3,3) & (4,2) & (4,3) \\ \hline
		{[}0,0.001)  & 43417 & 37687 & 40554 & 34629 \\ \hline
		{[}0.001,1)  & 4893  & 8767  & 7008  & 11607 \\ \hline
		{[}1,1.5)    & 1345  & 3177  & 2165  & 3577  \\ \hline
		{[}1.5,2)    & 345   & 369   & 273   & 187   \\ \hline
		{[}2,2.5)    & 0     & 0     & 0     & 0     \\ \hline
		{[}2.5,2.75) & 0     & 0     & 0     & 0     \\ \hline
		{[}2.75,3)   & 0     & 0     & 0     & 0     \\ \hline
		{[}3,3.1)    & 0     & 0     & 0     & 0     \\ \hline
		{[}3.1,3.17) & 0     & 0     & 0     & 0     \\ \hline
	\end{tabular}
	\caption{Histogram of certified randomness for 50000 expressions from Family 2 at noise level $0.99$.}
	\label{tab:Histogram_Family2_099}
\end{table}

\subsection{Robustness of the best randomly generated expressions}
\label{sec:RobustnessRandomBell}

We now analyze in more detail the robustness of the strongest randomly generated Bell expressions from Family~1 and Family~2. Their formulae are provided in App.~\ref{app:Bell_formulae}.

For each family and for each scenario $(3,2)$, $(3,3)$, $(4,2)$ and $(4,3)$, we selected the Bell expressions that certified the largest amount of min-entropy at noise levels $v=0.99$ and $v=0.99999$. In total, this procedure yielded six representative expressions for Family~1 and six for Family~2. We then evaluated their certified randomness as a continuous function of the noise parameter, both from the spot setting and when averaging over all inputs.

\begin{figure}[h!]
    \centering
    \includegraphics[width=0.99\linewidth]{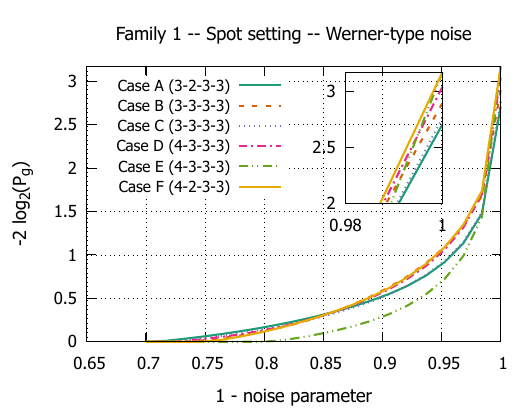}
    \caption{Certified global randomness for six selected Bell expressions from Family~1 as a function of the noise parameter $v$ under the Werner-type noise model of Eq.~\eqref{eq:werner_noise}. The inset highlights the region with more than two bits of certified randomness.}
    \label{fig:Family1_prob_noise}
\end{figure}

\begin{figure}[h!]
    \centering
    \includegraphics[width=0.99\linewidth]{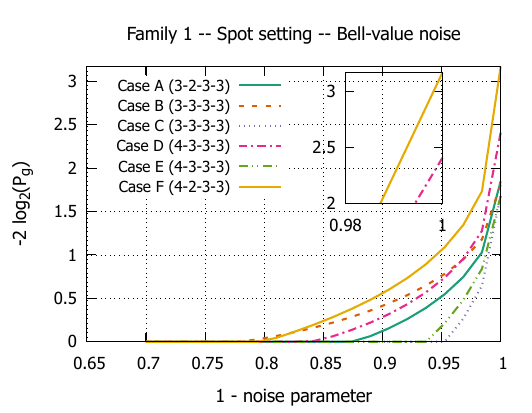}
    \caption{Certified global randomness for six selected Bell expressions from Family~1 as a function of the noisy Bell value $I_v$ defined in Eq.~\eqref{eq:noise_ineq}. The inset highlights the region with more than two bits of certified randomness.}
    \label{fig:Family1_bell_noise}
\end{figure}

Figure~\ref{fig:Family1_prob_noise} presents the certified global randomness under the Werner-type noise model of Eq.~\eqref{eq:werner_noise} for the selected expressions from Family~1 in the spot-setting case. The corresponding curves as a function of the noisy Bell value defined in Eq.~\eqref{eq:noise_ineq} are shown in Figure~\ref{fig:Family1_bell_noise}. The behavior closely resembles that observed previously for the manually selected inequalities: the best expressions in scenarios with four settings achieve the highest randomness in the near-ideal regime, while several three-setting scenarios exhibit improved robustness at moderate noise.

\begin{figure}[h!]
    \centering
    \includegraphics[width=0.99\linewidth]{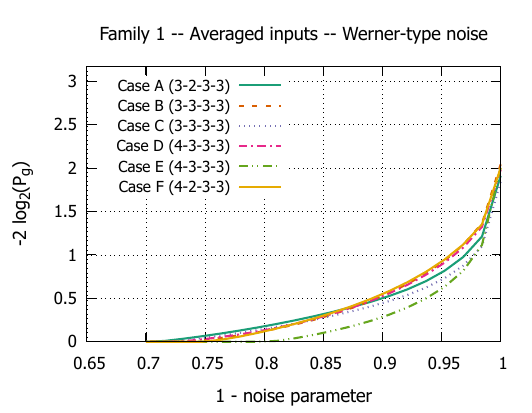}
    \caption{Certified global randomness from the spot setting for six selected Bell expressions from Family~1 as a function of the noise parameter $v$ under the Werner-type noise model of Eq.~\eqref{eq:werner_noise}.}
    \label{fig:Family1_prob_noise_avg}
\end{figure}

\begin{figure}
    \centering
    \includegraphics[width=0.99\linewidth]{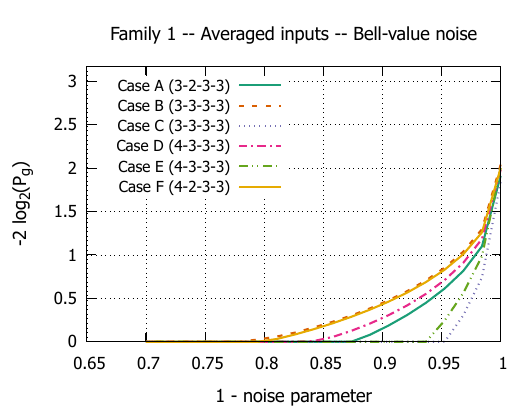}
    \caption{Certified global randomness from the spot setting for six selected Bell expressions from Family~1 as a function of the noisy Bell value $I_v$ defined in Eq.~\eqref{eq:noise_ineq}.}
    \label{fig:Family1_bell_noise_avg}
\end{figure}

We now turn to the case where the randomness is averaged over all inputs. Figures~\ref{fig:Family1_prob_noise_avg} and~\ref{fig:Family1_bell_noise_avg} present the corresponding certified randomness under the two noise parametrizations.

\begin{figure}[h!]
    \centering
    \includegraphics[width=0.99\linewidth]{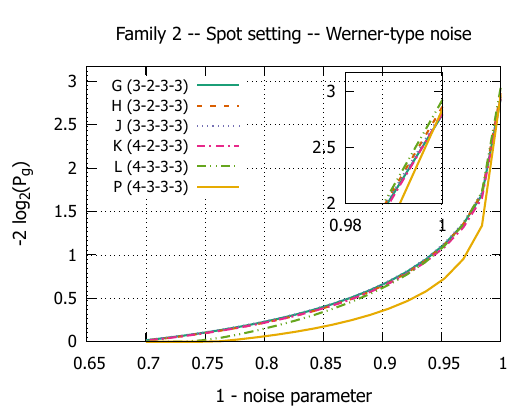}
    \caption{Certified global randomness for six selected Bell expressions from Family~2 as a function of the noise parameter $v$ under the Werner-type noise model of Eq.~\eqref{eq:werner_noise}. The inset highlights the region with more than two bits of certified randomness.}
    \label{fig:Family2_prob_noise}
\end{figure}

\begin{figure}[h!]
    \centering
    \includegraphics[width=0.99\linewidth]{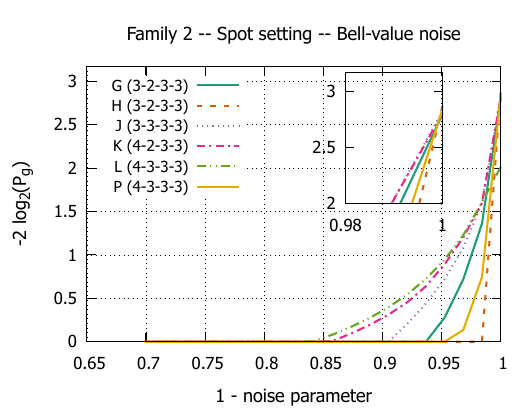}
    \caption{Certified global randomness for six selected Bell expressions from Family~2 as a function of the noisy Bell value $I_v$ defined in Eq.~\eqref{eq:noise_ineq}. The inset highlights the region with more than two bits of certified randomness.}
    \label{fig:Family2_bell_noise}
\end{figure}

\begin{figure}
    \centering
    \includegraphics[width=0.99\linewidth]{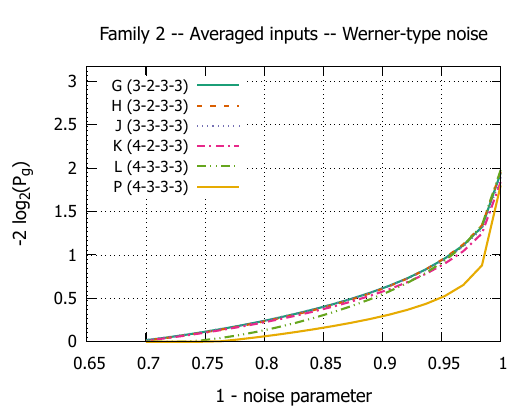}
    \caption{Averaged certified global randomness over all inputs for six selected Bell expressions from Family~2 as a function of the noise parameter $v$ under the Werner-type noise model of Eq.~\eqref{eq:werner_noise}.}
    \label{fig:Family2_prob_noise_avg}
\end{figure}

\begin{figure}
    \centering
    \includegraphics[width=0.99\linewidth]{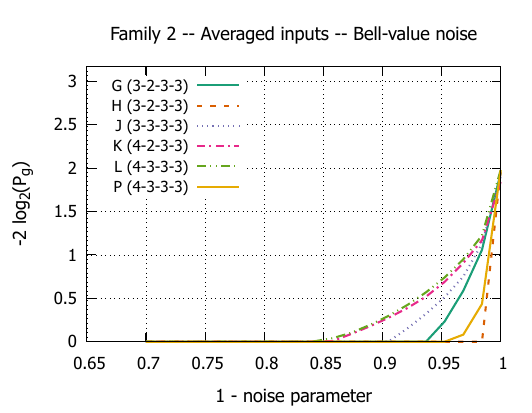}
    \caption{Averaged certified global randomness over all inputs for six selected Bell expressions from Family~2 as a function of the noisy Bell value $I_v$ defined in Eq.~\eqref{eq:noise_ineq}.}
    \label{fig:Family2_bell_noise_avg}
\end{figure}

Figures~\ref{fig:Family2_prob_noise} and~\ref{fig:Family2_bell_noise} show the analogous results for Family~2 in the spot-setting case. Although the maximal achievable randomness is generally lower than for the best instances from Family~1, the qualitative behavior is similar. The curves are smooth and monotonic in the noise parameter, and the certified randomness decreases gradually rather than abruptly, confirming robustness against noise.
The corresponding results when averaging over all inputs are shown in Figures~\ref{fig:Family2_prob_noise_avg} and~\ref{fig:Family2_bell_noise_avg}.

Overall, these findings show that strong robustness is not a rare feature of isolated, carefully engineered inequalities. Even when the Bell expressions are generated at random within structured families, the best-performing instances exhibit smooth degradation under noise and retain nonzero certified randomness well away from the ideal regime. This further supports the conclusion that high-quality device-independent randomness certificates arise generically within these families rather than as exceptional constructions.

\subsection{A near-maximal inequality with only two settings for one party}
\label{sec:InterestingF}

Among all randomly generated Bell expressions, one particular example from Family~1 stands out. It corresponds to the scenario $(4,2)$ with three outcomes per measurement and is denoted by ``F'' in the robustness plots of Sec.~\ref{sec:RobustnessRandomBell}. In this configuration one of the parties uses only two measurement settings, while the other uses four.

The Bell expression has the following form:
\begin{equation}
	\label{eq:F}
	\begin{aligned}
	    &P(0,0|0,1) + P(1,2|0,1) + P(2,1|0,1) \\
	    &+ P(0,2|0,2) + P(1,1|0,2) + P(2,0|0,2) \\
	    &+ P(0,2|0,3) + P(1,1|0,3) + P(2,0|0,3) \\
	    &+ P(0,1|1,0) + P(1,0|1,0) + P(2,2|1,0) \\
	    &+ P(0,1|1,1) + P(1,0|1,1) + P(2,2|1,1) \\
	    &+ P(0,1|1,2) + P(1,0|1,2) + P(2,2|1,2) \\
	    &+ P(0,2|1,3) + P(1,1|1,3) + P(2,0|1,3).
	\end{aligned}
\end{equation}

Despite its relatively simple structure, this inequality certifies an amount of global randomness extremely close to the theoretical maximum. It may be also shown, that up to relabelling of the settings and outcomes it is equivalent to the member of the $(2,4,3)$ family from~\cite{Barcelona}. Under the Bell-value noise model defined in Eq.~\eqref{eq:noise_ineq} it reaches $3.165021$ bits of min-entropy, while under the Werner-type (probability-level) noise model of Eq.~\eqref{eq:werner_noise} (equivalently Eq.~\eqref{eq:noise_distr}) it certifies $3.152592$ bits. Both values are remarkably close to the maximal value achievable in this scenario.

Several features make this example particularly noteworthy. First, the spot setting is $(0,0)$, so no additional measurement setting is required solely for randomness generation. Second, the inequality involves fewer measurement settings than other previously known Bell expressions that achieve near-maximal randomness.

Taken together, this example illustrates that near-optimal device-independent randomness certification does not require highly elaborate or heavily parameterized inequalities. Even comparatively simple Bell expressions arising from random generation within Family~1 can achieve performance essentially indistinguishable from the theoretical optimum, while maintaining a minimal measurement structure.

\subsection{Discussion}
\label{sec:Discussion}

Our results reveal two complementary aspects of device-independent randomness certification in multi-outcome scenarios.

First, moving beyond binary outcomes indeed allows one to certify larger amounts of global randomness. In the three-outcome case considered here, the maximal achievable min-entropy significantly exceeds the binary limit. Importantly, this enhancement is not confined to a handful of specially tailored Bell inequalities. The large-scale sampling within Family~1 and Family~2 shows that strong certificates are statistically abundant: among randomly generated expressions, a non-negligible fraction achieves performance comparable to the best previously known constructions.

Second, robustness is not an exceptional feature but rather a generic one. The histograms and robustness plots demonstrate that most nontrivial inequalities retain positive certified randomness under realistic noise levels. Even though the maximal value decreases with noise, the qualitative behavior is smooth and stable across families and scenarios. This indicates that the observed advantage of three-outcome measurements is not purely formal or restricted to the ideal regime.

A particularly illustrative example is the inequality in Eq.~\eqref{eq:F}, which certifies near-maximal global randomness while involving only two settings for one party. Its simple structure and minimal measurement requirements show that high performance does not necessarily demand highly complex Bell expressions. From a practical perspective, such inequalities improve the balance between certified output randomness and the input randomness required to select measurement settings in randomness expansion protocols~\cite{coudron2013infinite,miller2016robust}.

Taken together, these observations suggest that structured random generation is a powerful tool for discovering competitive Bell inequalities. Rather than relying exclusively on analytically constructed examples, one can explore constrained families numerically and identify expressions that offer favorable trade-offs between maximal entropy, robustness, and experimental simplicity.

\section{Conclusion}
\label{sec:Conclusion}

In this work, we have investigated the robustness of device-independent randomness generation in higher-dimensional bipartite systems. By analyzing a variety of Bell inequalities, including both well-known constructions such as SATWAP, CGLMP, Buhrman--Massar, Kaniewski \emph{et al.}, Meyer \emph{et al.} and Perito \emph{et al.}; as well as randomly generated inequalities in dimension $d=3$, we have quantified the amount of randomness that can be certified from the joint outputs of Alice and Bob. Our analysis shows that the inequality presented in \cite{Barcelona} is the one exhibiting more robustness to noise while achieving the two bit threshold imposed by the optimal projective qubit measurements.

Our results show that, in principle, maximal randomness can be achieved in ideal noiseless scenarios. However, even small amounts of noise lead to a decrease in the certifiable randomness, making robustness a central figure of merit. At the same time, we have demonstrated that multi-outcome scenarios allow one to surpass the randomness achievable with binary outputs, and that this advantage is not limited to carefully engineered inequalities. Many randomly generated Bell expressions within structured families certify substantial and robust amounts of min-entropy, with some simple examples, such as Eq.~\eqref{eq:F}, reaching values very close to the theoretical maximum.

These findings indicate that higher-dimensional correlations provide a genuine and practically accessible resource for device-independent randomness expansion. While the trade-off between certified randomness, robustness, and number of inputs remains crucial, our study shows that strong and experimentally attractive candidates can be identified systematically within well-chosen families of Bell expressions.

Taken together, our study provides a quantitative understanding of the trade-off between dimensionality and noise robustness in device-independent randomness generation, highlighting the practical limitations of high-dimensional protocols and the continued relevance of qubit-based approaches.

\section*{Acknowledgements}

RD acknowledges funding from the European Union’s Horizon Europe research and innovation programme under the Marie Skłodowska-Curie grant agreement No. 101081441. This work was supported by the European Union’s Horizon Europe research and innovation programme under grant agreement No 101080086/NeQST, the Government of Spain (Severo Ochoa CEX2019-000910-S and FUNQIP), Fundació Cellex, Fundació Mir-Puig, Generalitat de Catalunya (CERCA program), the EU (QSNP, 101114043, Quantera COMPUTE No 101017733) and the AXA Chair in Quantum Information Science.
The Center
for Quantum Enabled Computing project is carried out
within the International Research Agendas programme of
the Foundation for Polish Science co-financed by the Eu-
ropean Union under the European Funds for Smart Econ-
omy 2021-2027 (FENG).

\section*{Data and code availability}
The code used to perform the numerical optimizations and generate the figures presented in this work is available at \url{https://github.com/raffaeledavino/robust-three-outcome}.

\bibliography{References.bib}

@article{satwap,
  title = {Bell Inequalities Tailored to Maximally Entangled States},
  author = {Salavrakos, Alexia and Augusiak, Remigiusz and Tura, Jordi and Wittek, Peter and Ac\'{\i}n, Antonio and Pironio, Stefano},
  journal = {Phys. Rev. Lett.},
  volume = {119},
  issue = {4},
  pages = {040402},
  numpages = {6},
  year = {2017},
  month = {Jul},
  publisher = {American Physical Society},
  doi = {10.1103/PhysRevLett.119.040402},
  url = {https://link.aps.org/doi/10.1103/PhysRevLett.119.040402}
}

@article{nieto2014using,
  title={Using complete measurement statistics for optimal device-independent randomness evaluation},
  author={Nieto-Silleras, Olmo and Pironio, Stefano and Silman, Jonathan},
  journal={New Journal of Physics},
  volume={16},
  number={1},
  pages={013035},
  year={2014},
  publisher={IOP Publishing},
  doi = {10.1088/1367-2630/16/1/013035}
}

@article{clauser1969proposed,
  title={Proposed experiment to test local hidden-variable theories},
  author={Clauser, John F and Horne, Michael A and Shimony, Abner and Holt, Richard A},
  journal={Physical Review Letters},
  volume={23},
  number={15},
  pages={880},
  year={1969},
  publisher={APS},
  doi = {https://doi.org/10.1103/PhysRevLett.23.880}
}

@article{collins2002bell,
  title={Bell inequalities for arbitrarily high-dimensional systems},
  author={Collins, Daniel and Gisin, Nicolas and Linden, Noah and Massar, Serge and Popescu, Sandu},
  journal={Physical Review Letters},
  volume={88},
  number={4},
  pages={040404},
  year={2002},
  publisher={APS},
  doi ={https://doi.org/10.1103/PhysRevLett.88.040404}
}

@article{meyer2025robustly,
  title={Robustly self-testing all maximally entangled states in every finite dimension},
  author={Meyer, Uta Isabella and {\v{S}}upi{\'c}, Ivan and Grosshans, Fr{\'e}d{\'e}ric and Markham, Damian},
  journal={arXiv preprint arXiv:2508.01071},
  year={2025},
  doi = {https://doi.org/10.48550/arXiv.2508.01071}
}

@article{buhrman2005causality,
  title={Causality and Tsirelson's bounds},
  author={Buhrman, Harry and Massar, Serge},
  journal={Physical Review A—Atomic, Molecular, and Optical Physics},
  volume={72},
  number={5},
  pages={052103},
  year={2005},
  publisher={APS},
  doi = {https://doi.org/10.1103/PhysRevA.72.052103}
}

@article{navascues2007bounding,
  title={Bounding the set of quantum correlations},
  author={Navascu{\'e}s, Miguel and Pironio, Stefano and Ac{\'\i}n, Antonio},
  journal={Physical Review Letters},
  volume={98},
  number={1},
  pages={010401},
  year={2007},
  publisher={APS},
  doi = {https://doi.org/10.1103/PhysRevLett.98.010401}
}

@article{navascues2008convergent,
  title={A convergent hierarchy of semidefinite programs characterizing the set of quantum correlations},
  author={Navascu{\'e}s, Miguel and Pironio, Stefano and Ac{\'\i}n, Antonio},
  journal={New Journal of Physics},
  volume={10},
  number={7},
  pages={073013},
  year={2008},
  publisher={IOP Publishing},
  doi = {10.1088/1367-2630/10/7/073013}
}

@article{kaniewski2019maximal,
  title={Maximal nonlocality from maximal entanglement and mutually unbiased bases, and self-testing of two-qutrit quantum systems},
  author={Kaniewski, J{\k{e}}drzej and {\v{S}}upi{\'c}, Ivan and Tura, Jordi and Baccari, Flavio and Salavrakos, Alexia and Augusiak, Remigiusz},
  journal={Quantum},
  volume={3},
  pages={198},
  year={2019},
  publisher={Verein zur F{\"o}rderung des Open Access Publizierens in den Quantenwissenschaften},
  doi = {https://doi.org/10.22331/q-2019-10-24-198}
}

@article{mironowicz2013robustness,
  title={Robustness of quantum-randomness expansion protocols in the presence of noise},
  author={Mironowicz, Piotr and Paw{\l}owski, Marcin},
  journal={Physical Review A—Atomic, Molecular, and Optical Physics},
  volume={88},
  number={3},
  pages={032319},
  year={2013},
  publisher={APS},
  doi = {https://doi.org/10.1103/PhysRevA.88.032319}
}

@article{ma2016quantum,
  title={Quantum random number generation},
  author={Ma, Xiongfeng and Yuan, Xiao and Cao, Zhu and Qi, Bing and Zhang, Zhen},
  journal={npj Quantum Information},
  volume={2},
  number={1},
  pages={1--9},
  year={2016},
  publisher={Nature Publishing Group},
  doi = {https://doi.org/10.1038/npjqi.2016.21}
}

@article{pironio2010random,
  title={Random numbers certified by Bell’s theorem},
  author={Pironio, Stefano and Ac{\'\i}n, Antonio and Massar, Serge and de La Giroday, A Boyer and Matsukevich, Dzmitry N and Maunz, Peter and Olmschenk, Steven and Hayes, David and Luo, Lefroy and Manning, T Andrew and others},
  journal={Nature},
  volume={464},
  number={7291},
  pages={1021--1024},
  year={2010},
  publisher={Nature Publishing Group UK London},
  doi = {https://doi.org/10.1038/nature09008}
}

@article{mayers2003self,
  title={Self testing quantum apparatus},
  author={Mayers, Dominic and Yao, Andrew},
  journal={arXiv preprint quant-ph/0307205},
  year={2003},
  doi = {https://doi.org/10.48550/arXiv.quant-ph/0307205}
}

@article{liu2018device,
  title={Device-independent quantum random-number generation},
  author={Liu, Yang and Zhao, Qi and Li, Ming-Han and Guan, Jian-Yu and Zhang, Yanbao and Bai, Bing and Zhang, Weijun and Liu, Wen-Zhao and Wu, Cheng and Yuan, Xiao and others},
  journal={Nature},
  volume={562},
  number={7728},
  pages={548--551},
  year={2018},
  publisher={Nature Publishing Group UK London},
  doi = {https://doi.org/10.1038/s41586-018-0559-3}
}

@article{bancal2014more,
  title={More randomness from the same data},
  author={Bancal, Jean-Daniel and Sheridan, Lana and Scarani, Valerio},
  journal={New Journal of Physics},
  volume={16},
  number={3},
  pages={033011},
  year={2014},
  publisher={IOP Publishing},
  doi = {10.1088/1367-2630/16/3/033011}
}

@article{miller2017universal,
  title={Universal security for randomness expansion from the spot-checking protocol},
  author={Miller, Carl A and Shi, Yaoyun},
  journal={SIAM Journal on Computing},
  volume={46},
  number={4},
  pages={1304--1335},
  year={2017},
  publisher={SIAM},
  doi = {https://doi.org/10.1137/15M1044333}
}

@article{shalm2021device,
  title={Device-independent randomness expansion with entangled photons},
  author={Shalm, Lynden K and Zhang, Yanbao and Bienfang, Joshua C and Schlager, Collin and Stevens, Martin J and Mazurek, Michael D and Abell{\'a}n, Carlos and Amaya, Waldimar and Mitchell, Morgan W and Alhejji, Mohammad A and others},
  journal={Nature Physics},
  volume={17},
  number={4},
  pages={452--456},
  year={2021},
  publisher={Nature Publishing Group UK London},
  doi = {https://doi.org/10.1038/s41567-020-01153-4}
}

@article{liu2021device,
  title={Device-independent randomness expansion against quantum side information},
  author={Liu, Wen-Zhao and Li, Ming-Han and Ragy, Sammy and Zhao, Si-Ran and Bai, Bing and Liu, Yang and Brown, Peter J and Zhang, Jun and Colbeck, Roger and Fan, Jingyun and others},
  journal={Nature Physics},
  volume={17},
  number={4},
  pages={448--451},
  year={2021},
  publisher={Nature Publishing Group UK London},
  doi = {https://doi.org/10.1038/s41567-020-01147-2}
}

@article{brown2019framework,
  title={A framework for quantum-secure device-independent randomness expansion},
  author={Brown, Peter J and Ragy, Sammy and Colbeck, Roger},
  journal={IEEE Transactions on Information Theory},
  volume={66},
  number={5},
  pages={2964--2987},
  year={2019},
  publisher={IEEE},
  doi = {10.1109/TIT.2019.2960252}
}

@article{coudron2013infinite,
  title={Infinite randomness expansion and amplification with a constant number of devices},
  author={Coudron, Matthew and Yuen, Henry},
  journal={arXiv preprint arXiv:1310.6755},
  year={2013},
  doi = {https://doi.org/10.48550/arXiv.1310.6755}
}

@article{miller2016robust,
  title={Robust protocols for securely expanding randomness and distributing keys using untrusted quantum devices},
  author={Miller, Carl A and Shi, Yaoyun},
  journal={Journal of the ACM (JACM)},
  volume={63},
  number={4},
  pages={1--63},
  year={2016},
  publisher={ACM New York, NY, USA},
  doi = {https://doi.org/10.1145/2885493}
}

@article{Barcelona,
  title={Bell inequalities tailored to optimal global randomness certification},
  author={Perito, Ignacio and D'Avino, Raffaele and Jung, Micha{\l} and Mironowicz, Piotr and Ac{\'\i}n, Antonio and Augusiak, Remigiusz},
  journal={arXiv preprint arXiv:2606.21362},
  year={2026},
  doi = {https://doi.org/10.48550/arXiv.2606.21362}
}

@article{Mironowicz_2024,
doi = {10.1088/1751-8121/ad2b85},
url = {https://doi.org/10.1088/1751-8121/ad2b85},
year = {2024},
month = {apr},
publisher = {IOP Publishing},
volume = {57},
number = {16},
pages = {163002},
author = {Mironowicz, Piotr},
title = {Semi-definite programming and quantum information},
journal = {Journal of Physics A: Mathematical and Theoretical}
}

\clearpage
\onecolumngrid
\appendix

\begin{center}
\large{\textbf{\textsc{Supplementary Material}}}

\end{center}
\section{Formulae for randomized Bell expressions}
\label{app:Bell_formulae}

In this appendix we collect the explicit forms of the randomized Bell expressions discussed in the main text. 
For convenience, Table~\ref{tab:summary_randomized_Bell} summarizes their classification into families, the corresponding measurement scenario, the generating vector used in the construction procedure, and the label of the equation where the full expression is written explicitly.

The generating vectors uniquely specify the expressions within each family according to the construction rules introduced in Secs.~\ref{sec:Family1_definition} and~\ref{sec:Family2_definition}. The explicit probability expansions are provided below for completeness and reproducibility.

\begin{table}[h]
	\begin{tabular}{|l|l|l|l|l|}
		\hline
		Family & ID & scenario & vector                                                                                                  & Formula                       \\ \hline
		1      & A  & (3,2)    & {[}1, 1, 3, 1, 1{]}                                                                                     & \eqref{eq:A} \\ \hline
		1      & B  & (3,3)    & {[}3, 2, 3, 0, 1, 3, 3, 3{]}                                                                            & \eqref{eq:B} \\ \hline
		1      & C  & (3,3)    & {[}2, 2, 2, 5, 2, 3, 1, 4{]}                                                                            & \eqref{eq:C} \\ \hline
		1      & D  & (4,3)    & {[}3, 2, 1, 3, 0, 3, 0, 3, 2, 2, 5{]}                                                                   & \eqref{eq:D} \\ \hline
		1      & E  & (4,3)    & {[}3, 0, 1, 3, 2, 1, 1, 2, 1, 2, 2{]}                                                                   & \eqref{eq:E} \\ \hline
		1      & F  & (4,2)    & {[}1, 3, 3, 2, 2, 2, 3{]}                                                                               & \eqref{eq:F} \\ \hline \hline
		2      & G  & (3,2)    & {[}4, 3, 1, 4, 1, 3, 3, 1, 2, 4, 0, 1, 3, 2, 0{]}                                                       & \eqref{eq:G} \\ \hline
		2      & H  & (3,2)    & {[}1, 3, 4, 2, 4, 1, 4, 2, 1, 3, 4, 2, 4, 3, 1{]}                                                       & \eqref{eq:H} \\ \hline
		2      & J  & (3,3)    & {[}2, 4, 0, 3, 4, 2, 0, 3, 3, 2, 1, 3, 0, 1, 4, 0, 1, 3, 3, 1, 1, 1, 1, 2{]}                            & \eqref{eq:J} \\ \hline
		2      & K  & (4,2)    & {[}2, 3, 0, 2, 0, 4, 3, 0, 2, 2, 2, 2, 1, 1, 1, 3, 3, 2, 1, 4, 2{]}                                     & \eqref{eq:K} \\ \hline
		2      & L  & (4,3)    & {[}4, 3, 2, 3, 2, 3, 4, 3, 2, 3, 0, 2, 1, 1, 2, 4, 1, 0, 1, 4, 4, 1, 2, 0, 3, 3, 3, 3, 0, 3, 3, 0, 3{]} & \eqref{eq:L} \\ \hline
		2      & P  & (4,3)    & {[}3, 0, 4, 4, 0, 3, 1, 1, 3, 1, 3, 2, 1, 2, 3, 1, 3, 4, 2, 4, 1, 3, 2, 4, 2, 0, 3, 4, 4, 3, 3, 3, 4{]} & \eqref{eq:P} \\ \hline
	\end{tabular}
	\caption{Summary of selected randomized Bell expressions.}
	\label{tab:summary_randomized_Bell}
\end{table}

We now provide the explicit expressions in terms of conditional probabilities. 
All Bell expressions are written in the standard form as linear combinations of probabilities $P(a,b|x,y)$, where $a,b \in \{0,1,2\}$ denote outcomes and $x,y$ denote measurement settings according to the scenario specified in Table~\ref{tab:summary_randomized_Bell}. Positive and negative coefficients indicate contributions that increase or decrease the Bell value, respectively.

We begin with the Bell expressions belonging to Family~1.

Expression~A corresponds to Family~1 in the scenario $(3,2)$:
\begin{equation}
	\label{eq:A}
	\begin{aligned}
		&P(0,0|0,1) + P(1,2|0,1) + P(2,1|0,1) + P(0,0|1,0) + P(1,2|1,0) + P(2,1|1,0) + P(0,2|1,1) + P(1,1|1,1) \\
		&\qquad + P(2,0|1,1) + P(0,0|2,0) + P(1,2|2,0) + P(2,1|2,0) + P(0,0|2,1) + P(1,2|2,1) + P(2,1|2,1)
	\end{aligned}
\end{equation}
Expression~B belongs to Family~1 in the scenario $(3,3)$:
\begin{equation}
	\label{eq:B}
	\begin{aligned}
		&P(0,2|0,1) + P(1,1|0,1) + P(2,0|0,1) + P(0,1|0,2) + P(1,0|0,2) + P(2,2|0,2) + P(0,2|1,0) + P(1,1|1,0) \\
		&\qquad + P(2,0|1,0) + P(0,0|1,2) + P(1,2|1,2) + P(2,1|1,2) + P(0,2|2,0) + P(1,1|2,0) + P(2,0|2,0) \\
		&\qquad + P(0,2|2,1) + P(1,1|2,1) + P(2,0|2,1) + P(0,2|2,2) + P(1,1|2,2) + P(2,0|2,2)
	\end{aligned}
\end{equation}
Expression~C is another representative of Family~1 in the scenario $(3,3)$:
\begin{equation}
	\label{eq:C}
	\begin{aligned}
		&P(0,1|0,1) + P(1,0|0,1) + P(2,2|0,1) + P(0,1|0,2) + P(1,0|0,2) + P(2,2|0,2) + P(0,1|1,0) + P(1,0|1,0) \\
		&\qquad + P(2,2|1,0) - P(0,1|1,1) - P(1,0|1,1) - P(2,2|1,1) + P(0,1|1,2) + P(1,0|1,2) + P(2,2|1,2) \\
		&\qquad + P(0,2|2,0) + P(1,1|2,0) + P(2,0|2,0) + P(0,0|2,1) + P(1,2|2,1) + P(2,1|2,1) - P(0,0|2,2) \\
		&\qquad - P(1,2|2,2) - P(2,1|2,2)
	\end{aligned}
\end{equation}
Expression~D corresponds to Family~1 in the scenario $(4,3)$:
\begin{equation}
	\label{eq:D}
	\begin{aligned}
		&P(0,2|0,1) + P(1,1|0,1) + P(2,0|0,1) + P(0,1|0,2) + P(1,0|0,2) + P(2,2|0,2) + P(0,0|1,0) + P(1,2|1,0) \\
		&\qquad + P(2,1|1,0) + P(0,2|1,1) + P(1,1|1,1) + P(2,0|1,1) + P(0,2|2,0) + P(1,1|2,0) + P(2,0|2,0) \\
		&\qquad + P(0,2|2,2) + P(1,1|2,2) + P(2,0|2,2) + P(0,1|3,0) + P(1,0|3,0) + P(2,2|3,0) + P(0,1|3,1) \\
		&\qquad + P(1,0|3,1) + P(2,2|3,1) - P(0,1|3,2) - P(1,0|3,2) - P(2,2|3,2)
	\end{aligned}
\end{equation}
Expression~E is obtained within Family~1 for the scenario $(4,3)$:
\begin{equation}
	\label{eq:E}
	\begin{aligned}
		&P(0,2|0,1) + P(1,1|0,1) + P(2,0|0,1) + P(0,0|1,0) + P(1,2|1,0) + P(2,1|1,0) + P(0,2|1,1) + P(1,1|1,1) \\
		&\qquad + P(2,0|1,1) + P(0,1|1,2) + P(1,0|1,2) + P(2,2|1,2) + P(0,0|2,0) + P(1,2|2,0) + P(2,1|2,0) \\
		&\qquad + P(0,0|2,1) + P(1,2|2,1) + P(2,1|2,1) + P(0,1|2,2) + P(1,0|2,2) + P(2,2|2,2) + P(0,0|3,0) \\
		&\qquad + P(1,2|3,0) + P(2,1|3,0) + P(0,1|3,1) + P(1,0|3,1) + P(2,2|3,1) + P(0,1|3,2) + P(1,0|3,2) \\
		&\qquad + P(2,2|3,2)
	\end{aligned}
\end{equation}
Expression~F, discussed in Sec.~\ref{sec:InterestingF}, belongs to Family~1 in the scenario $(4,2)$, and is given in~\eqref{eq:F} in the main text.

We now turn to the Bell expressions constructed within Family~2.
Expression~G corresponds to Family~2 in the scenario $(3,2)$:
\begin{equation}
	\label{eq:G}
	\begin{aligned}
		&-2 P(0,0|0,1) - P(0,1|0,1) + P(0,2|0,1) - P(1,0|0,1) + P(1,1|0,1) - 2 P(1,2|0,1) + P(2,0|0,1) \\
		&\qquad - 2 P(2,1|0,1) - P(2,2|0,1) - 2 P(0,0|1,0) + P(0,1|1,0) - P(0,2|1,0) + P(1,0|1,0) - P(1,1|1,0) \\
		&\qquad - 2 P(1,2|1,0) - P(2,0|1,0) - 2 P(2,1|1,0) + P(2,2|1,0) - P(0,0|1,1) + P(0,1|1,1) \\
		&\qquad + 2 P(0,2|1,1) + P(1,0|1,1) + 2 P(1,1|1,1) - P(1,2|1,1) + 2 P(2,0|1,1) - P(2,1|1,1) \\
		&\qquad + P(2,2|1,1) - 2 P(0,0|2,0) + P(0,2|2,0) + P(1,1|2,0) - 2 P(1,2|2,0) + P(2,0|2,0) \\
		&\qquad - 2 P(2,1|2,0) - P(0,0|2,1) + 2 P(0,1|2,1) + 2 P(1,0|2,1) - P(1,2|2,1) - P(2,1|2,1) \\
		&\qquad + 2 P(2,2|2,1)
	\end{aligned}
\end{equation}
Expression~H is another example from Family~2 in the scenario $(3,2)$:
\begin{equation}
	\label{eq:H}
	\begin{aligned}
		&P(0,0|0,1) - P(0,1|0,1) - 2 P(0,2|0,1) - P(1,0|0,1) - 2 P(1,1|0,1) + P(1,2|0,1) - 2 P(2,0|0,1) \\
		&\qquad + P(2,1|0,1) - P(2,2|0,1) + 2 P(0,0|1,0) - 2 P(0,1|1,0) + P(0,2|1,0) - 2 P(1,0|1,0) \\
		&\qquad + P(1,1|1,0) + 2 P(1,2|1,0) + P(2,0|1,0) + 2 P(2,1|1,0) - 2 P(2,2|1,0) - 2 P(0,0|1,1) \\
		&\qquad + 2 P(0,1|1,1) + P(0,2|1,1) + 2 P(1,0|1,1) + P(1,1|1,1) - 2 P(1,2|1,1) + P(2,0|1,1) \\
		&\qquad - 2 P(2,1|1,1) + 2 P(2,2|1,1) - P(0,0|2,0) - 2 P(0,1|2,0) + 2 P(0,2|2,0) - 2 P(1,0|2,0) \\
		&\qquad + 2 P(1,1|2,0) - P(1,2|2,0) + 2 P(2,0|2,0) - P(2,1|2,0) - 2 P(2,2|2,0) - 2 P(0,0|2,1) \\
		&\qquad - P(0,1|2,1) + P(0,2|2,1) - P(1,0|2,1) + P(1,1|2,1) - 2 P(1,2|2,1) + P(2,0|2,1) - 2 P(2,1|2,1) \\
		&\qquad - P(2,2|2,1)
	\end{aligned}
\end{equation}
Expression~J belongs to Family~2 in the scenario $(3,3)$:
\begin{equation}
	\label{eq:J}
	\begin{aligned}
		&2 P(0,0|0,1) - 2 P(0,1|0,1) - 2 P(1,0|0,1) + 2 P(1,2|0,1) + 2 P(2,1|0,1) - 2 P(2,2|0,1) - P(0,0|0,2) \\
		&\qquad - 2 P(0,1|0,2) + 2 P(0,2|0,2) - 2 P(1,0|0,2) + 2 P(1,1|0,2) - P(1,2|0,2) + 2 P(2,0|0,2) - P(2,1|0,2) \\
		&\qquad - 2 P(2,2|0,2) - P(0,1|1,0) - P(0,2|1,0) - P(1,0|1,0) - P(1,1|1,0) - P(2,0|1,0) - P(2,2|1,0) + 2 P(0,0|1,1) \\
		&\qquad + P(0,1|1,1) - P(0,2|1,1) + P(1,0|1,1) - P(1,1|1,1) + 2 P(1,2|1,1) - P(2,0|1,1) + 2 P(2,1|1,1) + P(2,2|1,1) \\
		&\qquad + P(0,1|1,2) - 2 P(0,2|1,2) + P(1,0|1,2) - 2 P(1,1|1,2) - 2 P(2,0|1,2) + P(2,2|1,2) + P(0,1|2,0) \\
		&\qquad - P(0,2|2,0) + P(1,0|2,0) - P(1,1|2,0) - P(2,0|2,0) + P(2,2|2,0) - P(0,0|2,1) + P(0,1|2,1) \\
		&\qquad + P(0,2|2,1) + P(1,0|2,1) + P(1,1|2,1) - P(1,2|2,1) + P(2,0|2,1) - P(2,1|2,1) + P(2,2|2,1) \\
		&\qquad + P(0,0|2,2) + P(0,1|2,2) + 2 P(0,2|2,2) + P(1,0|2,2) + 2 P(1,1|2,2) + P(1,2|2,2) \\
		&\qquad + 2 P(2,0|2,2) + P(2,1|2,2) + P(2,2|2,2)
	\end{aligned}
\end{equation}
Expression~K corresponds to Family~2 in the scenario $(4,2)$:
\begin{equation}
	\label{eq:K}
	\begin{aligned}
		&2 P(0,0|0,1) - P(0,1|0,1) - P(1,0|0,1) + 2 P(1,2|0,1) + 2 P(2,1|0,1) - P(2,2|0,1) + 2 P(0,0|1,0) \\
		&\qquad - 2 P(0,2|1,0) - 2 P(1,1|1,0) + 2 P(1,2|1,0) - 2 P(2,0|1,0) + 2 P(2,1|1,0) - P(0,0|1,1) + 2 P(0,2|1,1) \\
		&\qquad + 2 P(1,1|1,1) - P(1,2|1,1) + 2 P(2,0|1,1) - P(2,1|1,1) + 2 P(0,0|2,0) + 2 P(0,1|2,0) + 2 P(0,2|2,0) \\
		&\qquad + 2 P(1,0|2,0) + 2 P(1,1|2,0) + 2 P(1,2|2,0) + 2 P(2,0|2,0) + 2 P(2,1|2,0) + 2 P(2,2|2,0) + P(0,0|2,1) \\
		&\qquad + P(0,1|2,1) + P(0,2|2,1) + P(1,0|2,1) + P(1,1|2,1) + P(1,2|2,1) + P(2,0|2,1) + P(2,1|2,1) \\
		&\qquad + P(2,2|2,1) - P(0,0|3,0) - P(0,1|3,0) + 2 P(0,2|3,0) - P(1,0|3,0) + 2 P(1,1|3,0) - P(1,2|3,0) \\
		&\qquad + 2 P(2,0|3,0) - P(2,1|3,0) - P(2,2|3,0) + P(0,0|3,1) - 2 P(0,1|3,1) + 2 P(0,2|3,1) - 2 P(1,0|3,1) \\
		&\qquad + 2 P(1,1|3,1) + P(1,2|3,1) + 2 P(2,0|3,1) + P(2,1|3,1) - 2 P(2,2|3,1)
	\end{aligned}
\end{equation}
Expression~L is obtained within Family~2 for the scenario $(4,3)$:
\begin{equation}
	\label{eq:L}
	\begin{aligned}
		&-2 P(0,0|0,1) - P(0,1|0,1) + 2 P(0,2|0,1) - P(1,0|0,1) + 2 P(1,1|0,1) - 2 P(1,2|0,1) + 2 P(2,0|0,1) \\
		&\qquad - 2 P(2,1|0,1) - P(2,2|0,1) - P(0,0|0,2) + 2 P(0,1|0,2) - P(0,2|0,2) + 2 P(1,0|0,2) - P(1,1|0,2) \\
		&\qquad - P(1,2|0,2) - P(2,0|0,2) - P(2,1|0,2) + 2 P(2,2|0,2) - 2 P(0,0|1,0) - P(0,1|1,0) + 2 P(0,2|1,0) \\
		&\qquad - P(1,0|1,0) + 2 P(1,1|1,0) - 2 P(1,2|1,0) + 2 P(2,0|1,0) - 2 P(2,1|1,0) - P(2,2|1,0) - P(0,0|1,1) \\
		&\qquad + 2 P(0,2|1,1) + 2 P(1,1|1,1) - P(1,2|1,1) + 2 P(2,0|1,1) - P(2,1|1,1) + P(0,0|1,2) + P(0,1|1,2) \\
		&\qquad + 2 P(0,2|1,2) + P(1,0|1,2) + 2 P(1,1|1,2) + P(1,2|1,2) + 2 P(2,0|1,2) + P(2,1|1,2) + P(2,2|1,2) \\
		&\qquad - 2 P(0,0|2,0) + P(0,1|2,0) + P(1,0|2,0) - 2 P(1,2|2,0) - 2 P(2,1|2,0) + P(2,2|2,0) + P(0,0|2,1) \\
		&\qquad - 2 P(0,1|2,1) - 2 P(0,2|2,1) - 2 P(1,0|2,1) - 2 P(1,1|2,1) + P(1,2|2,1) - 2 P(2,0|2,1) \\
		&\qquad + P(2,1|2,1) - 2 P(2,2|2,1) + P(0,0|2,2) + 2 P(0,1|2,2) + 2 P(1,0|2,2) + P(1,2|2,2) + P(2,1|2,2) \\
		&\qquad + 2 P(2,2|2,2) - P(0,0|3,0) - P(0,1|3,0) - P(0,2|3,0) - P(1,0|3,0) - P(1,1|3,0) - P(1,2|3,0) \\
		&\qquad - P(2,0|3,0) - P(2,1|3,0) - P(2,2|3,0) - P(0,0|3,1) - P(0,2|3,1) - P(1,1|3,1) - P(1,2|3,1) \\
		&\qquad - P(2,0|3,1) - P(2,1|3,1) - P(0,0|3,2) - P(0,2|3,2) - P(1,1|3,2) - P(1,2|3,2) - P(2,0|3,2) \\
		&\qquad - P(2,1|3,2)
	\end{aligned}
\end{equation}
Finally, Expression~P corresponds to Family~2 in the scenario $(4,3)$:
\begin{equation}
	\label{eq:P}
	\begin{aligned}
		&-P(0,0|0,1) - 2 P(0,2|0,1) - 2 P(1,1|0,1) - P(1,2|0,1) - 2 P(2,0|0,1) - P(2,1|0,1) - 2 P(0,0|0,2) \\
		&\qquad - P(0,2|0,2) - P(1,1|0,2) - 2 P(1,2|0,2) - P(2,0|0,2) - 2 P(2,1|0,2) + P(0,0|1,0) + P(0,1|1,0) \\
		&\qquad - P(0,2|1,0) + P(1,0|1,0) - P(1,1|1,0) + P(1,2|1,0) - P(2,0|1,0) + P(2,1|1,0) + P(2,2|1,0) \\
		&\qquad + P(0,0|1,1) - P(0,1|1,1) + 2 P(0,2|1,1) - P(1,0|1,1) + 2 P(1,1|1,1) + P(1,2|1,1) + 2 P(2,0|1,1) \\
		&\qquad + P(2,1|1,1) - P(2,2|1,1) + P(0,0|1,2) + 2 P(0,1|1,2) - P(0,2|1,2) + 2 P(1,0|1,2) - P(1,1|1,2) \\
		&\qquad + P(1,2|1,2) - P(2,0|1,2) + P(2,1|1,2) + 2 P(2,2|1,2) + P(0,0|2,0) - P(0,1|2,0) - 2 P(0,2|2,0) \\
		&\qquad - P(1,0|2,0) - 2 P(1,1|2,0) + P(1,2|2,0) - 2 P(2,0|2,0) + P(2,1|2,0) - P(2,2|2,0) + 2 P(0,0|2,1) \\
		&\qquad - 2 P(0,1|2,1) + P(0,2|2,1) - 2 P(1,0|2,1) + P(1,1|2,1) + 2 P(1,2|2,1) + P(2,0|2,1) \\
		&\qquad + 2 P(2,1|2,1) - 2 P(2,2|2,1) - P(0,0|2,2) + 2 P(0,1|2,2) - 2 P(0,2|2,2) + 2 P(1,0|2,2) \\
		&\qquad - 2 P(1,1|2,2) - P(1,2|2,2) - 2 P(2,0|2,2) - P(2,1|2,2) + 2 P(2,2|2,2) + 2 P(0,0|3,0) \\
		&\qquad - P(0,2|3,0) - P(1,1|3,0) + 2 P(1,2|3,0) - P(2,0|3,0) + 2 P(2,1|3,0) - 2 P(0,0|3,1) \\
		&\qquad - 2 P(0,1|3,1) - P(0,2|3,1) - 2 P(1,0|3,1) - P(1,1|3,1) - 2 P(1,2|3,1) - P(2,0|3,1) \\
		&\qquad - 2 P(2,1|3,1) - 2 P(2,2|3,1) - P(0,0|3,2) - P(0,1|3,2) - 2 P(0,2|3,2) - P(1,0|3,2) \\
		&\qquad - 2 P(1,1|3,2) - P(1,2|3,2) - 2 P(2,0|3,2) - P(2,1|3,2) - P(2,2|3,2)
	\end{aligned}
\end{equation}

\end{document}